\documentclass[journal,twoside,web]{ieeecolor}
\usepackage{etoolbox}
\makeatletter
\@ifundefined{color@begingroup}%
{\let\color@begingroup\relax
\let\color@endgroup\relax}{}%
\def\fix@ieeecolor@hbox#1{%
\hbox{\color@begingroup#1\color@endgroup}}
\patchcmd\@makecaption{\hbox}{\fix@ieeecolor@hbox}{}{\FAILED}
\patchcmd\@makecaption{\hbox}{\fix@ieeecolor@hbox}{}{\FAILED}
\usepackage{tmi}
\usepackage{cite}
\usepackage{amsmath,amssymb,amsfonts}
\usepackage{algorithm,algorithmic}
\usepackage{hyperref}
\hypersetup{hidelinks=true}
\usepackage{graphicx}
\usepackage{textcomp}
\usepackage{multirow}
\usepackage{color}
\usepackage{booktabs}
\usepackage{soul}
\soulregister\cite7
\soulregister\ref7
\def\BibTeX{{\rm B\kern-.05em{\sc i\kern-.025em b}\kern-.08em
    T\kern-.1667em\lower.7ex\hbox{E}\kern-.125emX}}
\begin{document}
\title{AVP-AP: Self-supervised Automatic View Positioning in 3D cardiac CT via Atlas Prompting}
\author{Xiaolin Fan, Yan Wang, Yingying Zhang, Mingkun Bao, Bosen Jia, Dong Lu, Yifan Gu, Jian Cheng, and Haogang Zhu
\thanks{This work was supported by the National Natural Science Foundation of China (No. 62406014 and U21A20523), Beijing Natural Science Foundation (No. 7244325, L222152, L242038, and 4252004), Start-up Funds of Hangzhou International Innovation Institute of Beihang University (No. 2024KQ045 and 2024KQ027), Peking University Third Hospital Fund for Interdisciplinary Research. (Corresponding authors: Yingying Zhang; Haogang Zhu; Jian Cheng)}
\thanks{Xiaolin Fan, Yan Wang, and Yifan Gu are with the School of Instrumentation and Optoelectronic Engineering, Beihang University, Beijing, 100191, China (e-mail:fanxiaolin.buaa@qq.com; wangyan9509@gmail.com; guyifan@buaa.edu.cn). }
\thanks{Mingkun Bao, Dong Lu, and Jian Cheng are with the School of Computer Science and Engineering, Beihang University, Beijing, 100191, China (e-mail: bravomikekilo@buaa.edu.cn; donglusx@gmail.com; jian\_cheng@buaa.edu.cn).}
\thanks{Bosen Jia is with the School of Biological Sciences, Victoria University of Wellington, Wellington, 6012, New Zealand, (e-mail: jiabose@myvuw.ac.nz).}
\thanks{Yingying Zhang is with the Hangzhou International Innovation Institute, Beihang University, Hangzhou, 311115, China, (e-mail: yingyingzhangbuaa@163.com).}
\thanks{Haogang Zhu is with the School of Computer Science and Engineering, Beihang University, Beijing, 100191, China, and also with the Hangzhou International Innovation Institute, Beihang University, Hangzhou, 311115, China, (e-mail: haogangzhu@buaa.edu.cn).}}

\maketitle
\begin{abstract}
Automatic view positioning is crucial for cardiac computed tomography (CT) examinations, including disease diagnosis and surgical planning. However, it is highly challenging due to individual variability and large 3D search space. Existing work needs labor-intensive and time-consuming manual annotations to train view-specific models, which are limited to predicting only a fixed set of planes. However, in real clinical scenarios, the challenge of positioning semantic 2D slices with any orientation into varying coordinate space in arbitrary 3D volume remains unsolved. We thus introduce a novel framework, AVP-AP, the first to use Atlas Prompting for self-supervised Automatic View Positioning in the 3D CT volume. Specifically, this paper first proposes an atlas prompting method, which generates a 3D canonical atlas and trains a network to map slices into their corresponding positions in the atlas space via a self-supervised manner. Then, guided by atlas prompts corresponding to the given query images in a reference CT, we identify the coarse positions of slices in the target CT volume using rigid transformation between the 3D atlas and target CT volume, effectively reducing the search space. Finally, we refine the coarse positions by maximizing the similarity between the predicted slices and the query images in the feature space of a given foundation model. Our framework is flexible and efficient compared to other methods, outperforming other methods by 19.8\% average structural similarity (SSIM) in arbitrary view positioning and achieving 9\% SSIM in two-chamber view compared to four radiologists. Meanwhile, experiments on a public dataset validate our framework's generalizability.
\end{abstract}

\begin{IEEEkeywords}
Automatic view positioning, Atlas prompting, Cardiac computed tomography, Foundation model, Self-supervised learning
\end{IEEEkeywords}

% 1 introduction
\section{Introduction}
\label{sec:introduction}
\begin{figure}[!t]
\centerline{\includegraphics[width=\columnwidth]{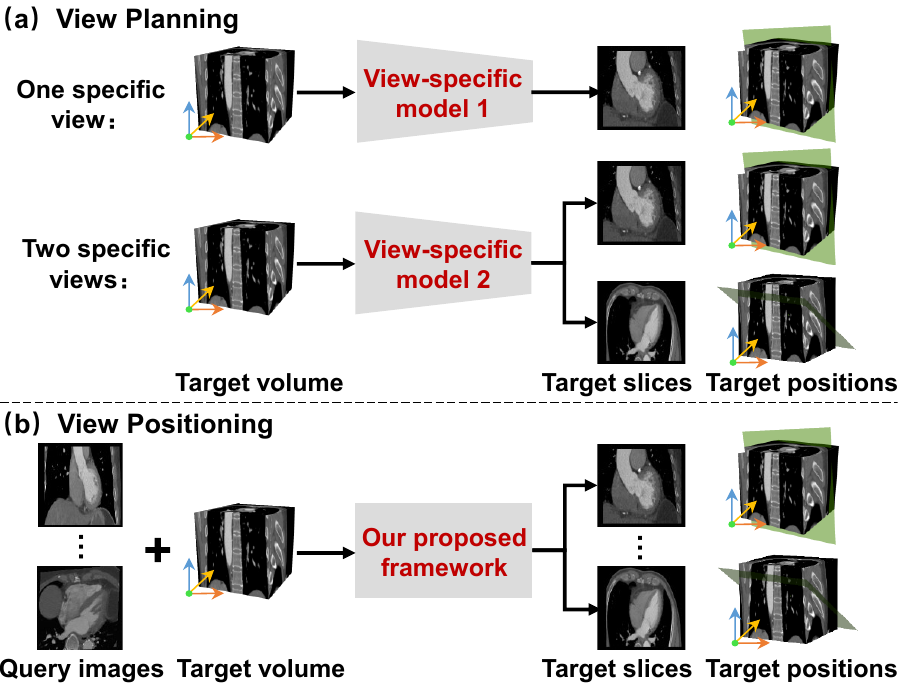}}
\caption{\textbf{An overview of view planning and view positioning.} (a) View planning: given a target volume, the view-specific model is trained to predict limited target 2D slices. When specific slices differ, the model needs to be retrained. (b) View positioning: for a target volume, given arbitrary query images in a reference CT, our framework can identify the most similar target 2D slices and their corresponding positions without the need for retraining.}
\label{fig:task}
\end{figure}

\IEEEPARstart{C}{ardiovascular} diseases (CVDs) are the leading cause of global mortality, resulting in up to 17.9 million deaths annually, representing 32\% of the total global deaths \cite{arnett20192019,luo2024global}. The use of high-definition cardiac images obtained through CT scans \cite{schuijf2022ct} enables clinicians to diagnose CVDs based on each patient's personalized anatomical structure, and strategically plan surgical procedures and treatments \cite{opolski2016ct,denner2024leveraging}.
For example, identifying standard views such as the long-axis four-chamber (4C) and arbitrary valve leaflet-related views from each CT volume can diagnose rheumatic mitral disease, and leveraging historical data to search for similar slices can assist better surgical planning \cite{wang2024scoring}. Therefore, using a prompt to position arbitrary 2D slices of interest in the target CT volume is of great significance. However, many current efforts in the analysis of cardiac CT images focus more on the segmentation of cardiac structures \cite{xie2022unsupervised,aromiwura2023artificial,lin2023deformable,miao2023sc}. Much less attention has been paid to automatic view positioning in 3D CT volume space (i.e., given a prompt, obtaining the 2D slices of interest and their corresponding positions in the target CT volume). Moreover, solving this problem in clinical practice faces many challenges. First, the views of cardiac CT are customized based on specific cardiac anatomical structures, which vary greatly between individuals, posing a challenge in establishing a unified coordinate space for view positioning. Second, mapping arbitrary 2D semantic slices into the large 3D search space exhibits high space and time complexity, which makes it challenging to implement in clinical practice efficiently. Hence, there is an urgent need for advanced techniques to address these challenges for automatic view positioning.

Currently, a few works have attempted automatic view planning for cardiac CT images, which primarily parameterizes the slice's position as three landmarks and further plans limited standard slices by identifying landmarks. We refer to this method as the view-specific method. For instance, Lu et al. \cite{lu2011automatic} segment the cardiac structures to obtain anatomical landmarks and calculate the standard views. Yuan et al.~\cite{yuan20222} initially evaluate the probability distribution of landmarks and further regress the positions of landmarks to achieve automatic view planning. However, these methods rely on segmenting anatomical structures or detecting landmarks to compute standard cardiac views. The prompt or prerequisite for these methods is obtaining prior knowledge of the entire 3D image for view detection. This involves manually annotating anatomical structures, which is tedious and time-consuming. Besides, the current methods can only retrieve specific views and are unable to query arbitrary slices, which is unsuited for certain clinical scenarios.

Another related work is rigid slice-to-volume registration, which aims to find the arbitrary slice's rigid transformation in 3D CT volumes (rigid transformation includes rotation and translation without altering the shape of images \cite{yuan2024learning,cao2023optical}). Among them, the optimization-based algorithms can map arbitrary 2D slices to different coordinate spaces in 3D CT volumes. For example, Porchetto et al. \cite{porchetto2017rigid} introduce a method to estimate the rigid transformation between slices and CT volumes using Markov random fields. However, these methods struggle with individual variability and often require a well-initialized position in large 3D space, complicating their utility in real-world scenarios. Also, the learning-based approaches only focus on aligning 2D slices of arbitrary orientation into a 3D fixed (or common) CT volume space \cite{ferrante2017slice}. For example, Hou et al. \cite{hou20183,hou2017predicting} propose a learning-based rigid slice-to-volume registration method for fetal brain images, which automatically learns the mapping function that aligns arbitrary slices into a canonical coordinate. Nevertheless, existing learning-based methods only map 2D slices to a unique coordinate of a fixed volume, making them unsuitable for addressing arbitrary coordinate spaces in different volumetric images. In addition, current methods have only been applied to static organs like the brain, while their suitability for dynamic organs, such as the heart, remains unexplored.

To address these limitations, in this paper, we propose a novel framework, AVP-AP, for self-supervised Automatic View Positioning in target 3D CT volume via Atlas Prompting. In detail, we first unify a coordinate system space and generate a 3D canonical atlas by a given sequence of 3D CT volumes. Then, we resample paired slice-position data from all CT volumes co-aligned to the atlas space and train a network to map any 2D slice into its corresponding position in the atlas space by a self-supervised manner. Next, guided by the atlas prompts (atlas and positions of query images in the atlas space, with query images being arbitrary slices of interest in a reference CT volume, see Fig.~\ref{fig:task}(b)), coarse positions in the target CT volume are determined by rigid transformations between the atlas and target CT volume. Finally, we refine the coarse positions by maximizing the similarity between the predicted slices and query images in the feature space of a given foundation model. As we use tailored atlas prompts for different input query images to locate target planes, the requirement for any manual annotations is eliminated. Additionally, we can standardize the implementation of the slice positioning workflow in arbitrary 3D cardiac CT volumes based on arbitrarily oriented image input, rather than a fixed plane. Specifically, our contributions are mainly summarized as follows: 

\begin{itemize}
  \item We propose a paradigm for standardized and personalized view positioning, designed for free-plane localization in various clinical tasks involving dynamic organs. Unlike the view-specific method, our approach can handle arbitrary slices without the need for retraining.
  \item We design an atlas prompting method that generates a 3D canonical atlas for cardiac CT, and implements automatic mappings from 2D slices with arbitrary orientation into the 3D atlas space.
  \item We propose an atlas prompting-based positioning framework AVP-AP in arbitrary target 3D CT volumes, which is guided by the atlas prompts to determine the most similar 2D slices and their positions in the 3D CT scans through a coarse-to-fine process.
  \item Our framework is flexible and efficient compared to other methods, outperforming other methods by 19.8\% average SSIM in arbitrary slice positioning, while achieving an average 9\% improvement in SSIM for the 2C view over four radiologists. Finally, experiments on a public dataset verify the generalization ability of our framework.
\end{itemize}

\begin{figure*}[!t]
\centering{\includegraphics[width=\textwidth]{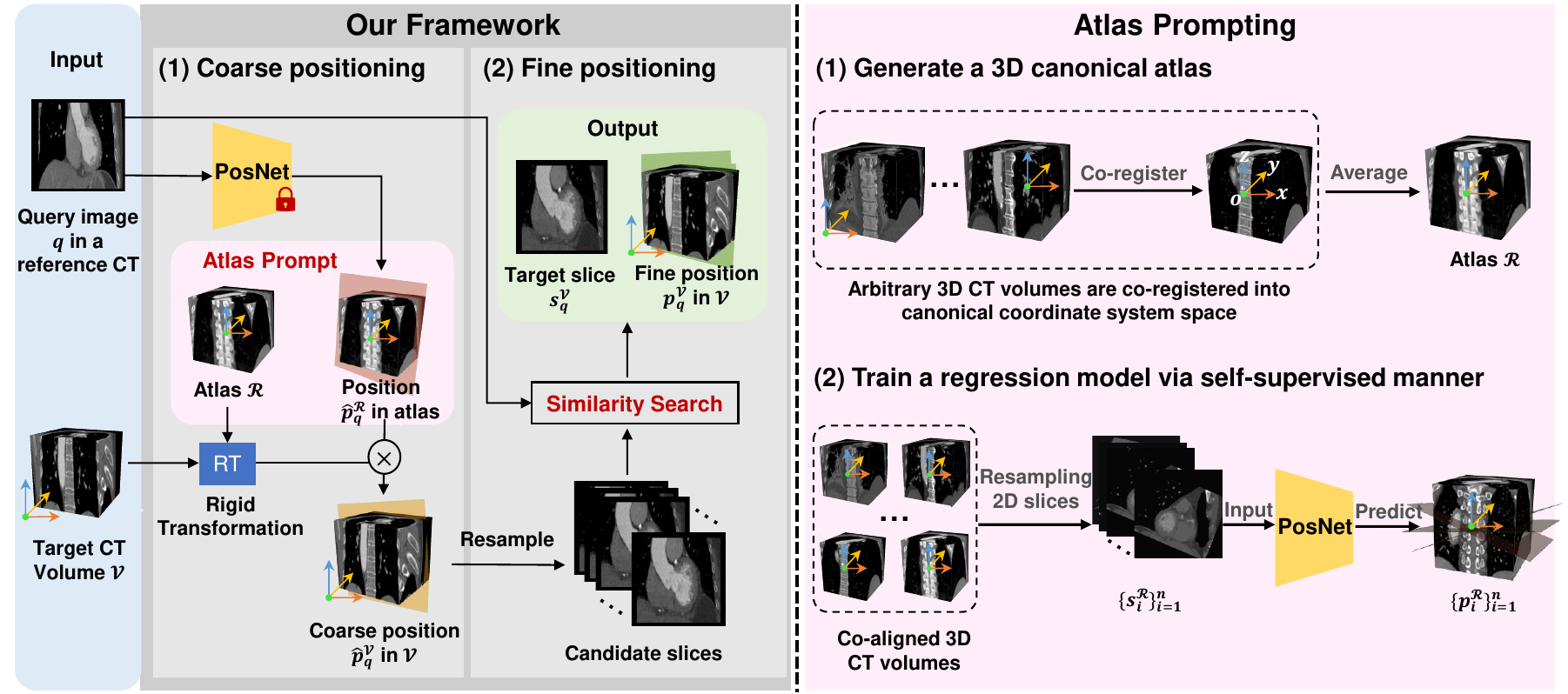}}
\caption{\textbf{Our framework, AVP-AP, for self-supervised Automatic View Positioning in 3D cardiac CT volume via Atlas Prompting.} For a target CT volume $\mathcal{V}$, given a query image $q$ in a reference CT, we first obtain the position $\hat{p}_q^\mathcal{R}$ of the query image in the atlas space using PosNet. Then, the atlas $\mathcal{R}$ and the position $\hat{p}_q^\mathcal{R}$ form the atlas prompt. Next, guided by the atlas prompt, the coarse position $\hat{p}_q^\mathcal{V}$ of the query image in the target CT volume is identified using rigid transformation between the atlas space and target CT volume. Finally, by maximizing the similarity between the predicted slices and the query images in the feature space of a given foundation model, the target slice $s_q^\mathcal{V}$ and its fine position $p_q^\mathcal{V}$ in the target CT volume is determined. For the atlas prompting method, it first unifies a coordinate system space and generates a 3D canonical atlas $\mathcal{R}$ with a given group of 3D CT volumes. Then, 2D image slices $\{s_i^{\mathcal{R}}\}_{i=1}^n$ resampled from all co-aligned 3D CT volumes are utilized to train a regression model PosNet to predict the positions $\{p_i^{\mathcal{R}}\}_{i=1}^n$ in atlas $\mathcal{R}$ via self-supervised learning.}\label{fig:framework}
\end{figure*}

% 2 Related Work
\section{Related Work}
\label{sec:relatedwork}
\subsection{Automatic View Planning}
Several studies have explored automatic view planning in cardiac volumetric images, which aims to acquire limited standard views by identifying three landmarks of slices. For example, Lu et al. \cite{lu2011automatic} segment the cardiac structures to obtain anatomical landmarks and calculate the standard views. Nunez-Garcia et al. \cite{nunez2021automatic} achieve view planning from the axial view to the short-axis plane by automatically segmenting the cardiac structure to obtain surface meshes for identifying the shape features. Le et al. \cite{le2017computationally} propose a novel method to generate cardiac long-axis and short-axis planes, which first computes a bounding box containing the heart and then locates landmarks within the box. Blansit et al. \cite{blansit2019deep} introduce a deep learning-based method to localize cardiac landmarks for visualizing standard views. Yuan et al. \cite{yuan20222} initially evaluate the probability distribution of landmarks and further regress the positions of landmarks to achieve automatic view planning. Next, Wei et al. \cite{wei2021training} introduce an innovative system to extract the spatial relationships between source and target views, and regress heatmaps defined by intersection lines to perform view planning. However, these approaches are based on segmenting anatomical structures or detecting landmarks to compute standard cardiac views. These methods require prior knowledge of the entire 3D image as a prompt or prerequisite for accurate view detection. This process involves manually annotating anatomical structures, which is laborious and time-consuming. In addition, existing methods are tailored to retrieving specific planes rather than any slices.

\subsection{Slice-to-Volume Registration}
Slice-to-volume registration is the process of aligning the slices (corresponding to arbitrary planes) into a unique coordinate of a fixed 3D volume \cite{ferrante2017slice}. Similar to our study is the rigid slice-to-volume registration, which is mainly divided into optimization-based and learning-based approaches. Porchetto et al. \cite{porchetto2017rigid} propose an optimization-based method to calculate the rigid transformation using Markov random fields. For learning-based methods, Hou et al. \cite{hou20183,hou2017predicting} propose a rigid slice-to-volume registration method for fetal brain images, which automatically learns the mapping that aligns arbitrary slices into a canonical coordinate, i.e., a volumetric atlas. Chen et al. \cite{chen2019real} develop a multi-label classification strategy for rigid slice-to-volume registration in adult brain images, estimating the 3D rigid transformation from 2D slices to 3D volume in common space. Mohseni Salehi et al. \cite{salehi2018real} introduce a novel geodesic loss to register arbitrarily oriented slices into a canonical fetal brain atlas, which broadens the capture range and enhances the performance of rigid 2D/3D registration. Zheng et al. \cite{zheng2024deep} propose an innovative rigid slice-to-volume registration method based on the differentiable sampling strategy, implementing the mapping of slices to the same subject in adult brain images. Khawaled et al. \cite{khawaled2024self} implement a slice-to-volume model designed to align 2D brain slices with a 3D fixed volume. Pei et al. \cite{pei2020anatomy} design a new multi-task learning framework aimed at simultaneously learning transformation parameters and brain segmentation maps for each input slice, for leveraging brain anatomical information to guide the mapping from 2D slices into 3D common volumetric space in a two-stage manner. Xu et al. \cite{xu2022svort} introduce a slice-to-volume registration method by employing transformers, which enables the model to seek the mapping functions from multiple stacks of fetal slices to a standard 3D space. However, existing deep learning-based studies primarily focus on establishing the mapping from 2D slices to a 3D fixed or canonical coordinate. Nevertheless, the coordinates of directly scanned volumetric images are non-uniform, and thus these methods cannot solve automatic view positioning. Moreover, current deep learning-based rigid slice-to-volume registration methods mainly focus on relatively standardized brain images and have not been explored for dynamic cardiac images. 

\subsection{Atlas-Based Segmentation}
Atlas-based segmentation has made significant progress, and combining an atlas with cutting-edge deep learning techniques can effectively enhance segmentation tasks. For example, Ding et al. \cite{ding2020cross,ding2022cross} propose a novel framework that registers the labeled atlas to the target image and uses the labels fusion strategy to combine the converted atlas labels to generate the segmentation of the target image. Ghosh et al. \cite{ghosh2021multi} introduce a statistical method that incorporates the prior knowledge described by atlas into the Bayes inference to obtain better cardiac anatomical structure segmentation labels. Mamalakis et al. \cite{mamalakis2023artificial} develop a pipeline that joins traditional multi-atlas segmentation with deep learning-based segmentation to achieve robust and generalized cardiac region segmentation. Therefore, it is evident that introducing prior knowledge of the statistical atlas into the deep learning-based segmentation network can enhance the robustness and generalization ability of the segmentation results, especially in different patients or pathological conditions. Similarly, our work leverages the prior slice position information from the atlas prompt, and combines classical registration technique with feature embedding similarity search to finally obtain the optimal result.

% 3 Methodology
\section{Methodology}
\label{sec:methodology}

% 3.1 Problem Definition and Framework Overview
\subsection{Problem Formulation and Framework Overview}
\subsubsection{Problem Formulation}
For a target 3D CT volume $\mathcal{V}$, given a query image $q$ in a reference CT, automatic view positioning is to obtain the most similar target 2D slice $s_q^{\mathcal{V}}=\mathcal{V}(p_q^{\mathcal{V}})$ at the corresponding position $p_q^{\mathcal{V}}$ of $\mathcal{V}$:
\begin{equation}
    p_q^{\mathcal{V}} = \mathop{\arg\max}\limits_{p}M(q, s), \, s = \mathcal{V}(p),p \in SE(3).
\end{equation}
where $M$ represents the image similarity term, $q$ is an arbitrary 2D slice of interest in a reference 3D CT volume and the slice $s$ is the 2D image mapped from the position $p$ in $\mathcal{V}$. The position $p = \{r,t\}$ is parameterized as a rigid transformation, i.e., rotation $r$ and translation $t$ from the initial plane (the initial plane is the $x$-$y$ plane in our unified coordinate system space). In detail, $p \in SE(3)$ lies within the special Euclidean group in 3D space \cite{hashim2021nonlinear}, $r \in SO(3)$ is in the special orthogonal group in 3D space \cite{fujun2023state}, and $t \in R^3$ is the translation in 3D space. As shown in Fig. \ref{fig:task}(b), for a target 3D CT volume $\mathcal{V}$, given arbitrary query images in a reference CT, we attain the most similar 2D slices and their corresponding positions in the target 3D CT volume $\mathcal{V}$.

\subsubsection{Framework Overview}
As shown in Fig. \ref{fig:framework}, we propose a novel two-stage framework, AVP-AP, to solve automatic view positioning in arbitrary target 3D cardiac CT volumes. In the \textbf{coarse} stage, the coarse position $\hat{p}_q^{\mathcal{V}}$ of the query image $q$ in the target 3D CT volume $\mathcal{V}$ is determined by the atlas prompt (i.e., atlas $\mathcal{R}$ and the predicted position $\hat{p}_q^\mathcal{R}$ of the query image $q$ in atlas $\mathcal{R}$) generated through our designed atlas prompting technique and rigid transformation between the atlas and target CT volume. During the \textbf{fine} stage, by multi-iteration resampling strategy at the coarse position $\hat{p}_q^{\mathcal{V}}$ in the target 3D CT volume $\mathcal{V}$ and employ image feature similarity search based on foundation model, we accurately identify the most similar 2D slice $s_q^\mathcal{V}$ and its corresponding position $p_q^\mathcal{V}$ in the target 3D CT volume $\mathcal{V}$. 

% 3.2 Atlas Prompting
\subsection{Atlas Prompting}
To achieve arbitrary view positioning, a query image needs to be introduced. However, owing to individual differences, the large 3D search space, and non-uniform coordinate spaces, direct view positioning within the target 3D cardiac CT volume presents significant complexity. Therefore, we further integrate a canonical atlas, and employ the atlas $\mathcal{R}$ and the position $\hat{p}_q^\mathcal{R}$ of the query image in the atlas space to "prompt" the approximate position of the query image $q$ in the target 3D CT volume $\mathcal{V}$. Here, the atlas $\mathcal{R}$ and the position $\hat{p}_q^\mathcal{R}$ of the query image $q$ in the atlas space are defined as the atlas prompt $\{ \mathcal{R}, \hat{p}_q^\mathcal{R} \}$. Moreover, our atlas prompting method can be applied to query images of arbitrary orientation, without the necessity for retraining. Below, we will elaborate on our atlas prompting method.

\subsubsection{Generate a 3D Canonical Atlas}
To address the challenges arising from individual variability and non-uniform coordinate system spaces, we generate a canonical atlas $\mathcal{R}$ from a given set of 3D CT volumes, serving as a standardized reference coordinate space for automatic view positioning in arbitrary CT volumes. As depicted in the Atlas Prompting part of Fig. \ref{fig:framework}, we first establish a unified coordinate system, where the origin is centered within the 3D CT volume, and the positive directions of the $x$, $y$, $z$ axes are represented by orange, yellow, and blue arrows, respectively. Then, all 3D CT volumes are co-registered into this canonical coordinate space. Finally, a multi-iteration averaging algorithm \cite{gholipour2017normative,wang2020allen} is employed to generate the 3D canonical atlas $\mathcal{R}$ with size $(L, L, L)$ from all co-registered 3D CT volumes.

\begin{figure}[t]
\centering
\begin{minipage}{0.3\linewidth}
    \vspace{9pt}
    \centerline{\includegraphics[width=\textwidth]{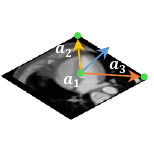}}
    \centerline{\footnotesize (a) Coordinate system}
\end{minipage}
\begin{minipage}{0.33\linewidth}
    \vspace{3pt}
    \centerline{\includegraphics[width=\textwidth]{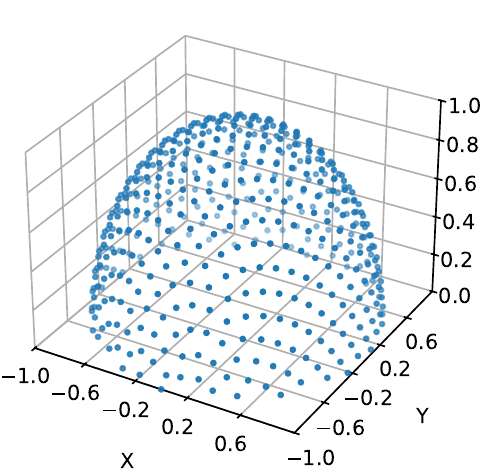}} 
    \centerline{\footnotesize (b) Normal sampling}
\end{minipage}
\begin{minipage}{0.33\linewidth}
    \vspace{3pt}
    \centerline{\includegraphics[width=\textwidth]{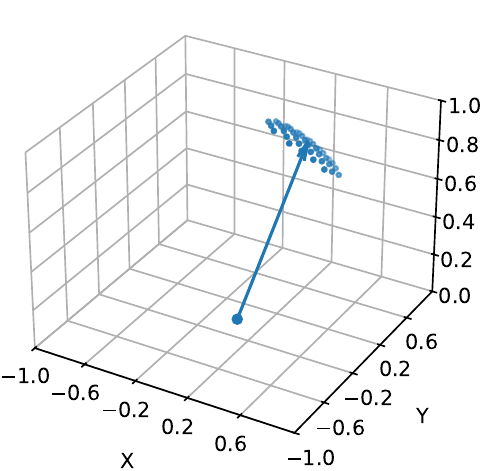}} 
    \centerline{\footnotesize (c) Positioning and searching}
\end{minipage}
\caption{\textbf{The coordinate system, normal sampling, and positioning and searching of slices.} (a) The coordinate system and Three-Point label representation of 2D slice self in 3D CT volume space. (b) Slice plane normals $\Omega$ w.r.t. the origin resampled by FSS. (c) Multi-iteration resampling normals w.r.t. the origin at the normal of the coarse position $\hat{p}_q^{\mathcal{V}}$ (i.e., blue arrow)}\label{fig:sampling}
\end{figure}

\subsubsection{Training a Regression Model}
After the canonical atlas $\mathcal{R}$ is generated, we aim to automatically and efficiently determine the position $\hat{p}_q^\mathcal{R}$ of a query image $q$ within the atlas space $\mathcal{R}$. To achieve this, all CT volumes are first aligned to the atlas space. Then, the paired slice-position data are resampled from the co-aligned CT volumes to train a regression model PosNet $\alpha(\cdot;\omega)$ for mapping arbitrary 2D slices into their corresponding positions in the atlas space. We provide a detailed explanation below.

First of all, we need to automatically resample sufficient and balanced slice-position pair data from all co-aligned 3D CT volumes. To generate balanced paired data, producing a lot of uniform rigid transformation parameters is essential. Here, we define a 3D rotation as a combination of the transformation from 3D vector $A$ to 3D vector $B$ and in-plane 2D rotation: $A$ is $(0,0,1)$, $B$ is the sampling normals of target 2D slices; in-plane 2D rotation is rotated around the z-axis, and it does not change the content of the slices. Among them, $B$ are generated by Fibonacci sphere sampling (FSS) \cite{marques2019extensible}, while in-plane rotations are selected uniformly within $(0, \pi)$. Each normal generated by FSS has approximately the same degree of separation from its adjacent normals. In detail, the sampling normals are figured out using polar coordinates $\rho({\phi}_i, cos^{-1}(z_i))$, where ${\phi}_i = 2\pi i/\Phi $ and $z_i = 1 - (2i+1)/n_o$, $i \in 0,1,2,\dots, n_o-1$. $\Phi = (\sqrt{5}+1)/2$ is the golden ratio and thus $\Phi^{-1}=\Phi-1$. For translation, we set it to start from the center of the CT volume and translate only along the normal vector of 3D rotations, with the translation range of $(-0.33L, 0.33L)$. By this resampling strategy, $n$ paired slice-position data $\{s_i^\mathcal{R},p_i^\mathcal{R}\}_{i=1}^n$ are generated. 

Then, we train a regression model PosNet $\alpha(\cdot;\omega)$ using the paired dataset $\{s_i^\mathcal{R},p_i^\mathcal{R}\}_{i=1}^n$. The PosNet $\alpha(\cdot;\omega)$ consists of the encoder module $\alpha_{enc}(\cdot;\omega_{enc})$ and the position prediction module $\alpha_{pos}(\cdot;\omega_{pos})$. Each 2D slice $s_i^{\mathcal{R}}$ is fed into the encoder module $\alpha_{enc}(\cdot;\omega_{enc})$ to generate a feature vector $f_i$:
\begin{equation}
\{f_1, \cdots, f_n \} = \{\alpha_{enc}(s_1^{\mathcal{R}};\omega_{enc}), \cdots,  \alpha_{enc}(s_n^{\mathcal{R}};\omega_{enc})\}.
\end{equation}
Following, we predict the positions $\{\hat{p}_i^\mathcal{R}\}_{i=1}^n$ of 2D slices $\{s_i^\mathcal{R}\}_{i=1}^n$ in the 3D atlas space $\mathcal{R}$ by feeding the feature vectors $\{f_i\}_{i=1}^n$ into the position prediction module $\alpha_{pos}(\cdot;\omega_{pos})$:
\begin{equation}
\{\hat{p}_1^\mathcal{R}, \cdots, \hat{p}_n^\mathcal{R}\} = \{\alpha_{pos}(f_1;\omega_{pos}), \cdots,  \alpha_{pos}(f_n;\omega_{pos})\}.  
\end{equation}

As defined by the problem formulation, the predicted position $\{\hat{p}_i^\mathcal{R}\}_{i=1}^n$ is parameterized by the rigid transformation, i.e., rotation and translation. Rigid transformation can be represented in multiple ways, such as Cartesian translation and rotation: rotation vectors, quaternions, rotation matrices, etc. Here, we use rotation vectors with the fewest three parameters and the Cartesian translation with three parameters, referring to this type of representation as the Rotvec-Cartesian label. Computing the loss function in this representation involves the sum of the geodesic distance (GD) \cite{salehi2018real} between 3D rotations and the weighted mean squared error (MSE) \cite{varshney2024optimizing} of 3D translations:
\begin{equation}
    Loss_{lc} =  GD(\hat{r}, r) + \lambda MSE(\hat{t}, t).
\end{equation}
where $\lambda$ is the weight of the MSE loss, $\hat{r}$ and $\hat{t}$ are predicted rotations and translations, $r$ and $t$ are ground truth one. $GD(\hat{r}, r)$ is the geodesic distance between the predicted rotations $\hat{r}$ and the ground truth rotations $r$, it can be written as follows:
\begin{equation}\label{eq:gd}
    GD(\hat{r}, r) = d(\hat{r},r) = cos^{-1} \left[ \frac{tr(\hat{r}^{T}r)-1}{2} \right].
\end{equation}
$MSE(\hat{t}, t)$ is the mean squared error between the predicted translations $\hat{t}$ and the ground truth translation $t$, which can be calculated as: 
\begin{equation}\label{eq:mse}
    MSE(\hat{t}, t) = {\| {\hat{t}} - t \|}^2_2.
\end{equation}
However, this combination of rotation in SO(3) and translation in the Cartesian system is not within the same evaluation range, which can lead to problems such as slow and non-smooth convergence (see Fig. \ref{fig:loss_line}(b)). 

In addition, Hou et al. \cite{hou2017predicting,hou20183} introduce a new Three-Point label to denote rotation and translation jointly. Three non-collinear points in 3D Euclidean space define a plane, with their arrangement determining the orientation. Moreover, three points can be placed anywhere on a 2D slice $s_i$, provided they are neither identical nor collinear. Here, we define $a_1$ as the center point, $a_2$ as the upper-left corner, and  $a_3$ as the upper-right corner of slice $s$ (see Fig. \ref{fig:sampling} (a)). In this way, $a_1(x_1,y_1,z_1)$, $a_2(x_2,y_2,z_2)$, $a_3(x_3,y_3,z_3)$ have a total of nine parameters, all of which are Cartesian. Thus, their combined optimization is balanced, and the loss function can be directly computed using mean squared error:
\begin{equation}
    Loss_{tp} = MSE(\hat{a}_{1}, a_{1}) + MSE(\hat{a}_{2}, a_{2}) + MSE(\hat{a}_{3}, a_{3}).
\end{equation}
where $\hat{a}_{1}, \hat{a}_{2}, \hat{a}_{3}$ are the predicted points, and $a_{1}, a_{2}, a_{3}$ are the ground truth points. The calculation of $MSE(\hat{a}_{1}, a_{1})$, $MSE(\hat{a}_{2}, a_{2})$, and $MSE(\hat{a}_{3}, a_{3})$ is same as in Equation (\ref{eq:mse}). Moreover, the experiments verify that both loss functions for these two representations of rigid transformations have their own advantages. For more details, see Section \ref{sec:loss}. 

\begin{algorithm}[t]
\caption{View Positioning.}\label{alg:alg1}
\renewcommand{\algorithmicrequire}{\textbf{Input:}}
\renewcommand{\algorithmicensure}{\textbf{Output:}}
\begin{algorithmic}[1]
\REQUIRE The query image $q$; The target 3D CT volume $\mathcal{V}$; \\
   The fixed bounding range $\gamma$; The normals $\Omega$ on the sphere. % input
\ENSURE The most similar slice $s_q^\mathcal{V}$ and position $p_q^\mathcal{V}$. % output
\STATE The \textbf{atlas prompt} $\{\mathcal{R}, \hat{p}_q^\mathcal{R} = \alpha(q;\omega)\}$
\STATE $RT \gets$ rigid transformation from $\mathcal{R}$ to $\mathcal{V}$ 
\STATE The \textbf{coarse position} $\hat{p}_q^{\mathcal{V}} =  RT( \hat{p}_q^\mathcal{R})$
\STATE Initialize \textbf{the current position} $p^{c} = \hat{p}_q^{\mathcal{V}}$, and \textbf{corresponding slice} $s^{c} \in \mathcal{V}(p^{c})$
\STATE The \textbf{feature} $f^c$=BiomedCLIP($s^{c}$), $f_q$=BiomedCLIP($q$)
\STATE The \textbf{baseline} similarity $sim^{c}$ = CSIM($f^c$, $f_q$)
\REPEAT
    \STATE $\{p_i^c\}_{i=1}^{m}$ = \textbf{Resample}($p^{c}$, $\gamma$, $\Omega$) 
    \STATE $\{f_i^c\}_{i=1}^{m}$ = BiomedCLIP($\mathcal{V}(\{p_i^c\}_{i=1}^{m})$)
        \STATE $p_i^{c*}$ = $\mathop{\max}\limits_{p_i^c}$ (CSIM($\{f_i^c\}_{i=1}^{m}$, $f_q$))
    \STATE $p^c$ = $p_i^{c*}$, $sim^c$ = CSIM($f_i^{c*}$, $f_q$)
\UNTIL{$sim^c$ \textbf{remains unchanged or starts to decrease}}
\RETURN the \textbf{most similar slice} $s_q^\mathcal{V}$ and \textbf{fine position} $p_q^\mathcal{V}$ corresponding to $f_q^{\mathcal{V}}$
\end{algorithmic}
\end{algorithm}

\subsection{View Positioning}
Given a query image $q$ in a reference CT, the atlas prompt $\{ \mathcal{R}, \hat{p}_q^\mathcal{R} \}$ corresponding to the query image $q$ is first obtained through the aforementioned atlas prompting technique. Then, the coarse position $\hat{p}_q^{\mathcal{V}}$ of the query image $q$ in the target 3D CT volume $\mathcal{V}$ is acquired via the guidance of the atlas prompt and rigid transformation between 3D atlas and target 3D CT volume. Finally, we finely determine the most similar 2D slice $s_q^\mathcal{V}$ and its corresponding position $p_q^\mathcal{V}$ in the target 3D CT volume $\mathcal{V}$, which employs a multi-iteration resampling strategy at the coarse position $\hat{p}_q^{\mathcal{V}}$ and uses the similarity metric of the feature embeddings extracted by the foundation model. Algorithm \ref{alg:alg1} describes the process of view positioning in detail.

\subsubsection{Coarse Positioning}
The query image $q$ is fed into the pre-trained model PosNet $\alpha(\cdot;\omega)$ to predict the position $\hat{p}_q^\mathcal{R}$ of the query image $q$ in the atlas space $\mathcal{R}$:
\begin{equation}
    \hat{p}_q^\mathcal{R} = \alpha(q;\omega).
\end{equation}
Then, the atlas $\mathcal{R}$ and the predicted position $\hat{p}_q^\mathcal{R}$ form the atlas prompt $\{\mathcal{R}, \hat{p}_q^\mathcal{R} \}$ for the query image $q$ (line 1).

For coarse positioning, the 3D atlas is rigidly registered into the target 3D CT volume space to obtain rigid transformation matrix $RT$ (the reason for using rigid registration here is that the position $p$ of the 2D slice $s$ is also expressed by rigid transformation). Next, $RT$ is applied into the predicted position $\hat{p}_q^\mathcal{R}$ to obtain the coarse position $\hat{p}_q^{\mathcal{V}}$ of the query image $q$ in the target 3D CT volume $\mathcal{V}$ (lines 2-3):
\begin{equation}
    \hat{p}_q^{\mathcal{V}} =  RT(\hat{p}_q^\mathcal{R}).
\end{equation}

\subsubsection{Fine Positioning}
After coarse positioning, we adopt a multi-iteration resampling strategy at the coarse position $\hat{p}_q^{\mathcal{V}}$ within the target 3D CT volume $\mathcal{V}$ for fine positioning. Benefiting from the strong anatomical feature extraction capability of the foundation model, we exploit the cosine similarity (CSIM) between the feature embeddings extracted by the foundation model as the similarity metric \cite{ye2019unsupervised}:
\begin{equation}
    \mathrm{CSIM} = \frac{f^c_i \cdot f^c_j}{\| f^c_i \| \|f^c_j\|}.
\end{equation}
where $f^c_i$ and $f^c_j$ are the feature vectors of the corresponding to images $s_i$ and $s_j$ extracted by the foundation model.
Moreover, according to the research on content-based image retrieval using foundation models proposed by Denner et al. \cite{denner2024leveraging}, BiomedCLIP \cite{zhang2023biomedclip} demonstrated superior performance in retrieving anatomical structures. Therefore, we use the features extracted by BiomedCLIP for similarity search to determine the most similar slice $s_q^\mathcal{V}$ and its corresponding position $p_q^\mathcal{V}$.

Specifically, a) we resample the corresponding slice $s^c$ at the coarse location $p^c = \hat{p}_q^{\mathcal{V}}$ and calculate the feature similarity $sim^c$ between this slice $s^c$ and the given query image $q$ as a baseline (lines 4-6); b) we set an initial fixed bounding range $\gamma$ around the position $p^c$, and resample slices within the bounding range to calculate the feature similarity between the resampled slices and the query slice to determine the position with the highest similarity (lines 8-10); c) repeat b) at the position $p^c$ obtained in the previous step until similarity remains unchanged or begins to decrease (lines 7-12). Finally, the most similar slice $s_q^\mathcal{V}$ and fine position $p_q^\mathcal{V}$ corresponding to feature embedding $f_q^\mathcal{V}$ are returned (line 13).

% 4 Experimental Setup
\section{Experiments and Results}
\label{sec:experiment}
\subsection{Datasets and Experimental Setup}
\subsubsection{Datasets}
Two datasets from different sources are collected. The first one is utilized to construct the atlas prompting and for internal testing of view positioning. The second one is used for external generalization validation of view positioning.

\textit{Dataset 1:} 
This dataset contains 80 reconstructed 3D cardiac CT volumes from 80 patients ranging in age from 29 to 72 years (mean=55.1, std=14.2) on the Siemens scanners. The origin CT data was acquired from the Medical Imaging Center of Anzhen Hospital\footnote{https://www.anzhen.org/Html/Departments/Main/Detail\_87.html}, and all subjects used in this study are with ethical committee approval. We use the 60 collected CT volumes to construct atlas prompting, i.e., generating a 3D canonical atlas and randomly dividing into training, validation, and test set in a ratio of 8:1:1 to train a regression model PosNet for predicting the position of the query image in the atlas space. Meanwhile, the remaining 20 CT volumes are used for internal testing of view positioning. 

\textit{Dataset 2:} This dataset includes 60 reconstructed 3D cardiac CT volumes. The data are acquired from a public database, Multi-Modality Whole Heart Segmentation (MM-WHS) Challenge 2017 Dataset\footnote{https://zmiclab.github.io/zxh/0/mmwhs/}. To match the number of CTs in internal testing, we randomly selected 20 CT volumes from this dataset for external generalization validation of view positioning.

\subsubsection{Implementation Details}
For atlas prompting, we respectively choose three different registration methods: Symmetric image Normalization registration (SyN) \cite{avants2008symmetric}, Affine registration (Affine) \cite{bhalerao2018construction}, and Rigid registration (Rigid) \cite{yuan2024learning} for atlas generation and alignment of all original CT volumes to atlas space. In detail, SyN is a deformable registration method. Affine includes rotation, translation, and scaling, while Rigid involves only rotation and translation. The size of the generated atlas $\mathcal{R}$ is 180 $\times$ 180 $\times$ 180 (i.e., $L$ = 180), and the isotropic resolution is 1.0 $\times$ 1.0 $\times$ 1.0 mm.

Then, to train the PosNet model in the atlas prompting, the paired slice-position data are generated by resampling from all co-aligned CT volumes using the above resampling strategy. The ground truth labels of the data have two representations: a Rotvec-Cartesian label and a Three-Point label.
In this way, three atlases and three types of paired data with two labels (a Rotvec-Cartesian label and a Three-Point label) are generated based on the three types of aligned 3D CT volumes. Afterward, the resampling parameters for the paired slice-position data are set. For each co-aligned 3D CT volume, 1500 rotations were applied, and translations were set from the center of the volume along the normal of rotation plane in 5 mm intervals from -60 mm to 60 mm. This resampling strategy accounts for approximately the middle 66.67\% of the entire CT volume. This involves resampling 36000 pairs of images and corresponding positions in each co-aligned 3D CT volume, resulting in a total of 2.16 million pairs of data across all co-aligned 3D CT volumes. Overall, this 2.16 million paired slice-position data is used to train the PosNet model for mapping any 2D slice into its corresponding position in the atlas space. % and then we can precisely locate the slice from coarse to fine in the target CT volumes.} 

Next, we select the appropriate network backbone. Since the main contribution of this paper is the AVP-AP framework rather than the backbone and our framework is independent of the backbone, we tested five classic backbones and selected the best-performing one. In detail, we chose four convolutional neural networks (CNN)-based backbones: ResNet18 \cite{he2016deep}, ResNet50 \cite{he2016deep}, MobineNetV2 \cite{sandler2018mobilenetv2}, VGGNet19 \cite{simonyan2014very}, and a transformer-based Swin-T \cite{liu2021swin} as the encoder module of PosNet $\alpha(\cdot;\omega)$ for ablation study. And, the position prediction module employs two fully connected layers \cite{long2015fully} to regress the positions of the 2D slices in the atlas space. In the training phase, input images are resized to 256 $\times$ 256, and randomly applied to brightness transformation data augmentation. The batch size is 256, and the AdamW optimizer \cite{loshchilov2017decoupled} with a weight decay of $10^{-3}$ for $100$ epochs is used. Further, we initialize the learning rate of $5 \times 10^{-4}$ for the CNN-based networks and $1 \times 10^{-4}$ for Swin-T, and adopt a learning rate adjustment strategy \cite{wang2022automatic} that decays by 0.1 times every 40 epochs.

Finally, algorithm \ref{alg:alg1} was implemented for internal testing and external generalization validation of view positioning. The fixed bounding range $\gamma$ is set to 6 by default. For the internal testing, we select long-axis 2-chamber (2C), long-axis 4-chamber (4C), transverse orthogonal view Y, and random view (RV$_1$) as input images, and then perform the view positioning algorithm on 20 target 3D CT volumes of \textit{Dataset 1}. Furthermore, we asked four radiologists to identify and label 2D slices in the above 20 target 3D CT volumes based on the given four-type query images. The annotation method first determines the 2C plane using fixed planning rules, followed by locating the 4C, and RV$_1$, finally relocates Y planes. Simultaneously, we record the time spent by radiologists annotating each slice. For another external generalization validation, we similarly choose 2C, 4C, Y, and RV$_1$ slices, and perform the view positioning algorithm on the 20 target CT volumes from \textit{Dataset 2}.
All experiments were implemented using Python \cite{python2021python} and PyTorch \cite{paszke2019pytorch} on a Nvidia Tesla V100 GPU.

\begin{table*}[t]
\centering
\caption{Internal testing: Mean values and standard deviation of image similarity between given query images and the results. NA: not applicable. The $\uparrow$ suggests bigger values being better, and vice versa. The best results are in bold.}\label{tab:our-com} 
\begin{tabular}{ccccccccc}
\toprule
 \multirow{2}{*}{Meth.} & \multicolumn{2}{c}{2C}      & \multicolumn{2}{c}{4C}      & \multicolumn{2}{c}{Y}    & \multicolumn{2}{c}{RV$_1$}     \\ \cmidrule{2-9}
           & SSIM $\uparrow$      & MSE $\downarrow$           & SSIM $\uparrow$      & MSE $\downarrow$          & SSIM $\uparrow$      & MSE $\downarrow$ & SSIM $\uparrow$      & MSE  $\downarrow$            \\ \cmidrule{1-9}
Opt-SVR \cite{porchetto2017rigid}  & 0.22 (0.09) & 4407.68 (512.2) & 0.32 (0.06) & 4027.82 (497.2) & 0.25 (0.06) & 4273.75 (502.4) & 0.29 (0.05) & 3976.20 (507.7)   \\
Lea-AVP \cite{nunez2021automatic} & 0.50 (0.03) & 1449.82 (328.9) & 0.54 (0.05) & 1495.82 (276.9) & NA & NA & NA & NA  \\
AVP-AP (ours) & \textbf{0.54} (0.07) & \textbf{1401.78} (326.7) & \textbf{0.55} (0.10) & \textbf{1390.63} (301.6) & \textbf{0.57} (0.08) & \textbf{1296.84} (354.4) & \textbf{0.56} (0.09) & \textbf{1217.56} (294.5)  \\
\bottomrule
\end{tabular}
\end{table*}

\subsubsection{Evaluation Metric}
For the evaluation of atlas prompting, we employ the following metrics: geodesic distance error (GD) and Euclidean distance error (ED) \cite{rass2024metricizing}.

GD is the geodesic distance between the predicted rotation $\hat{r}$ and the ground truth rotation $r$, which is calculated as shown in Equation (\ref{eq:gd}).

ED refers to the Euclidean distance between the predicted point (translation) $\hat{a}_i$ and the ground truth point (translation) $a_i$, which can be calculated as:
\begin{equation}
    \mathrm{ED} = {\| {\hat{a}}_i - {a}_i \|}_2, i \in (1, 2, 3).
\end{equation}

During the view positioning, we use standard image feature similarity metrics such as structural similarity (SSIM) \cite{bakurov2022structural}, and MSE.

SSIM represents the similarity of images in terms of luminance, contrast, and structure as perceived by the human visual system, with values ranging from 0 to 1. It can be written as:
\begin{equation}
    \mathrm{SSIM}(q,s)=\frac{(2\mu_q\mu_s+c_1)(2\sigma_{qs}+c_2)}{(\mu_q^2+\mu_s^2+c_1)(\sigma_q^2+\sigma_s^2+c_2)}.
\end{equation}
where $\mu_q$ and $\mu_s$ represent the mean pixel intensities of images $q$ and $s$ respectively; $\sigma_q^2$ and $\sigma_s^2$ are the variances of images $q$ and $s$; $\sigma_{qs}$ stands for the covariance of $q$ and $s$; $c_1$ and $c_2$ are two variables used to stabilize divisions with weak denominators.

MSE calculates the average of squared differences between corresponding pixel values in two images $q$ and $s$:
\begin{equation}
    \mathrm{MSE}(q,s)=\frac{1}{hw}\sum_i^{h-1}\sum_j^{w-1}\left(q(i,j)-s(i,j)\right)^2.
\end{equation}
where $q(i,j)$ and $s(i, j)$ are the pixel values at the corresponding $(i,j)$ positions of two images $q(i,j)$ and $s(i, j)$, respectively.

\begin{figure}[t]
\centering{\includegraphics[width=\columnwidth]{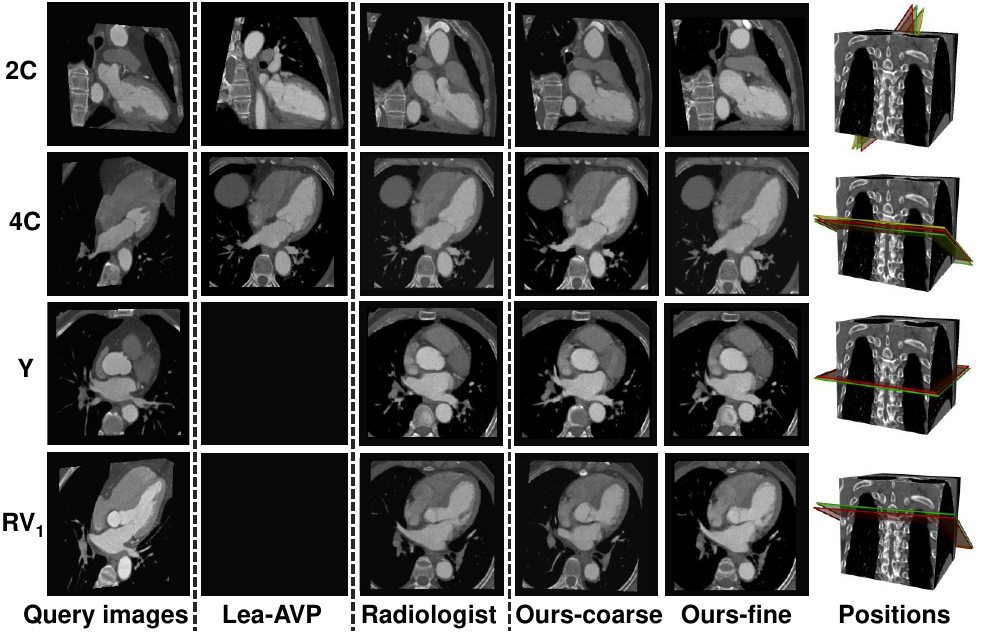}}
\caption{Internal testing: visualization of results from three different methods. The first column contains four query images: 2C, 4C, Y, and RV$_1$, not from the target CT volume. The second and third columns are the slices localized by Lea-AVP and the radiologist, while the fourth and fifth columns show the results of coarse and fine positioning by our method, with "Ours-coarse" representing the coarse positioning and "Ours-fine" indicating the fine positioning. The last column shows the comparison of positions. Green is our position, red is the result labeled by the radiologist, and yellow is the result located by Lea-AVP.}\label{fig:vis_res_an}
\end{figure}

\begin{table*}[t]
\centering
\caption{Internal testing: Mean values of image similarity quantitative metrics between given query images and the results labeled by radiologists (standard deviation in parentheses). Rad1 represents the first radiologist, the same as below. The $\uparrow$ suggests higher values being better, and vice versa. The best results are in bold.}\label{tab:simi_ourrad}
\begin{tabular}{cccccccccc}
\toprule
\multirow{2}{*}{Meth.} & \multicolumn{2}{c}{2C}  & \multicolumn{2}{c}{4C} & \multicolumn{2}{c}{Y} & \multicolumn{2}{c}{RV$_1$}   \\  \cmidrule{2-9}
                  & SSIM $\uparrow$ & MSE $\downarrow$ & SSIM $\uparrow$  & MSE $\downarrow$ & SSIM $\uparrow$ & MSE $\downarrow$ & SSIM $\uparrow$ & MSE $\downarrow$  \\ \cmidrule{1-9}
rad1              & 0.44 (0.06) & 1912.82 (567.2) & 0.52 (0.03) & 1711.45 (515.0) & 0.47 (0.04) & 1764.43 (494.3) & 0.49 (0.05) & 1672.15 (382.3)  \\
rad2              & 0.47 (0.06) & 1891.83 (572.6) & 0.51 (0.04) & 1598.31 (275.1) & 0.49 (0.06) & 1698.09 (437.0) & 0.49 (0.06) & 1674.47 (619.6)  \\
rad3              & 0.44 (0.05)  & 1908.17 (657.2) & 0.52 (0.04) & 1693.38 (572.1) & 0.50 (0.03) & 1623.87 (388.9) & 0.49 (0.06) & 1613.59 (324.9)  \\
rad4              & 0.45 (0.08) & 1831.45 (471.9) & 0.52 (0.04) & 1562.67 (476.6) & 0.48 (0.05) & 1639.58 (429.8) & 0.51 (0.07) & 1618.02 (609.4)  \\
ours              & \textbf{0.54} (0.07) & \textbf{1401.78} (326.7) & \textbf{0.55} (0.10)  & \textbf{1390.63} (301.6) & \textbf{0.57} (0.08) & \textbf{1296.84} (354.4) & \textbf{0.56} (0.09)  & \textbf{1217.56} (294.5)  \\ \bottomrule 
\end{tabular}
\end{table*}

\subsection{Comparison Experiments}
\label{sec:exis_real}
We first conduct internal testing, i.e., comparing our proposed framework AVP-AP with two existing methods, and then perform real-world validation with the identification results from four radiologists. Finally, external generalization validation is carried out.
\subsubsection{Internal Testing - Comparison with Existing Methods}
We select two typical comparison methods to validate the effectiveness of our framework AVP-AP in the 20 target CT volumes of \textit{Dataset 1}, which employs the atlas prompts with the best configuration in Section \ref{sec:ablation}. Table \ref{tab:our-com} presents the comparison results among the optimization-based method Opt-SVR \cite{porchetto2017rigid}, the learning-based method Lea-AVP \cite{nunez2021automatic}, and our proposed framework AVP-AP. We can observe that our framework AVP-AP outperforms Opt-SVR by a large margin, and achieves a 4\% improvement in SSIM of the 2C view compared to Lea-AVP. This reveals that iterative optimization methods struggle to accurately locate arbitrary 2D slices within a large 3D search space because it usually depends on a well-initialized position. In contrast, the learning-based methods can more precisely locate certain standard slices by using prior anatomical knowledge. However, the learning-based methods for automatic view planning, such as Lea-AVP, are unsuitable for locating arbitrary slices. As a result, the metrics of the Y and RV$_1$ views in Table \ref{tab:our-com} cannot be calculated, limiting their applicability. Fig. \ref{fig:vis_res_an} presents the visualization of the results from the other methods and our proposed framework AVP-AP. It can be seen that our results are better than Lea-AVP on 2C and comparable on 4C, while Lea-AVP cannot position the arbitrary Y and RV$_1$ slices. Hence, our framework can accurately position arbitrary 2D slices in arbitrary target 3D CT volumes through our designed atlas prompts, in comparison to existing methods. At the same time, it can be noted that our atlas prompting provides a well-initialized coarse position in the target CT volumes, significantly reducing the search space.

\begin{figure}[!t]
\centering
\begin{minipage}{0.24\linewidth}
    \vspace{3pt}
    \centerline{\includegraphics[width=\textwidth]{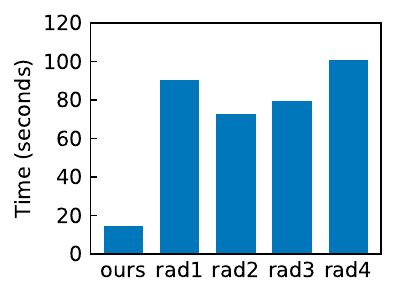}}
    \centerline{(a) 2C}
\end{minipage}
\begin{minipage}{0.24\linewidth}
    \vspace{3pt}
    \centerline{\includegraphics[width=\textwidth]{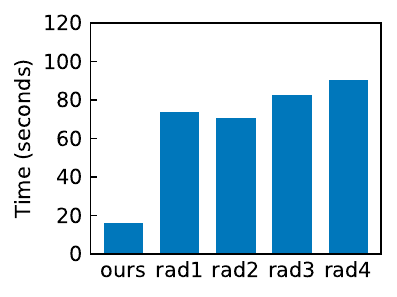}}
    \centerline{(b) 4C}
\end{minipage}
\begin{minipage}{0.24\linewidth}
    \centerline{\includegraphics[width=\textwidth]{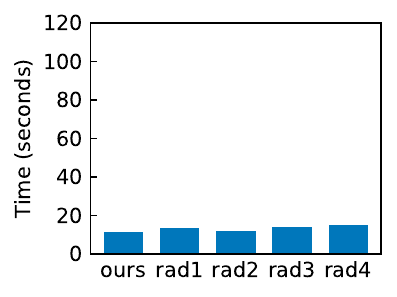}}
    \centerline{(c) Y }
\end{minipage}
\begin{minipage}{0.24\linewidth}
    \vspace{3pt}
    \centerline{\includegraphics[width=\textwidth]{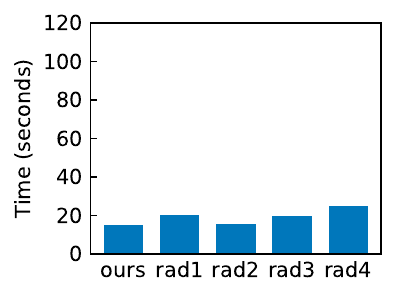}}
    \centerline{(d) RV$_1$}
\end{minipage}
\caption{Internal testing: comparison of positioning times between our framework AVP-AP and radiologists. The less time spent, the better. (a), (b), (c), and (d) represent the comparison of positioning time for four different types of query images: 2C, 4C, Y, and RV$_1$. Rad1 represents the first radiologist, the same as below.}
\label{fig:time}
\end{figure}

\begin{table*}[t]
\centering
\caption{External generalization validation: Mean values and standard deviation of image similarity between given query images and the results in generalization validation. NA: not applicable. The $\uparrow$ suggests bigger values being better, and vice versa. The best results are in bold.}\label{tab:our-mmwhs} 
\begin{tabular}{ccccccccc}
\toprule
 \multirow{2}{*}{Meth.} & \multicolumn{2}{c}{2C}      & \multicolumn{2}{c}{4C}   & \multicolumn{2}{c}{Y}   & \multicolumn{2}{c}{RV$_1$}         \\ \cmidrule{2-9}
           & SSIM $\uparrow$      & MSE $\downarrow$           & SSIM $\uparrow$      & MSE $\downarrow$           & SSIM $\uparrow$      & MSE  $\downarrow$   & SSIM $\uparrow$      & MSE $\downarrow$         \\ \cmidrule{1-9}
Opt-SVR \cite{porchetto2017rigid}  & 0.21 (0.10) & 4573.67 (624.1) & 0.29 (0.05) & 4187.26 (508.6) & 0.24 (0.06) & 4356.76 (498.1) & 0.23 (0.04) & 4067.76 (487.3)  \\
Lea-AVP \cite{nunez2021automatic} & 0.49 (0.08) & 1634.09 (389.5) & 0.52 (0.04) & 1598.45 (387.2) & NA & NA & NA & NA \\
AVP-AP (ours) & \textbf{0.54} (0.06) & \textbf{1456.37} (376.5) & \textbf{0.55} (0.08) & \textbf{1428.87} (359.2) & \textbf{0.56} (0.06) & \textbf{1438.65} (387.9) & \textbf{0.52} (0.08) & \textbf{1477.09} (360.5)  \\
\bottomrule
\end{tabular}
\end{table*}

\subsubsection{Internal Testing - Real-world Verification}
To verify the real-world clinical application ability of our framework AVP-AP, we compare our positioning results with those of four radiologists. As shown in Table \ref{tab:simi_ourrad}, we can see that our proposed method accomplishes an average improvement of 6.8\% in SSIM for arbitrary slice positioning compared to four radiologists. This proves that our approach achieves a better level of performance than clinical experts in view positioning from a large 3D CT volume space. As shown in Fig. \ref{fig:vis_res_an}, we can observe that the results of our AVP-AP in automatic view positioning are highly similar to the slices annotated by radiologists, especially on the 2C plane, where the images obtained by our proposed method are even closer to the query images, particularly regarding the structure of the left ventricle. This verifies the clinical applicability of our proposed framework AVP-AP in view positioning.

As shown in Fig. \ref{fig:time}, we can see that our method takes less time than all radiologists for both 2C, 4C, Y, and RV$_1$, demonstrating the efficiency of our approach in direct view positioning. However, for the RV$_1$ slices, we observed only a slight improvement in the positioning time of AVP-AP compared to that of the radiologists. By analyzing the results of Fig. \ref{fig:vis_res_an}, we found that the RV$_1$ slice is similar to and positioned close to the 4C slice, while the 2C slice differs significantly from the 4C slice and is located much farther away. Since the process of positioning the RV$_1$ slice by doctors is continuous—i.e., they position the RV$_1$ slice directly after the 4C view—radiologists have a smaller search space compared to AVP-AP, rather than starting from scratch. For the Y slice, the overall lower positioning time is because the orthogonal Y slice is localized directly in the transverse plane, without adjusting the rotation direction. As a result, the search space is greatly reduced. In conclusion, our framework AVP-AP is more effective for direct view positioning and is well-suited for flexible clinical applications. 

\begin{table*}[!t] %h-here, t-top, b-bottom, p-page
\centering
\caption{Mean error and standard deviation of PosNet positioning metrics for different registration methods and network architectures. Rotation (Rot.) error and Translation (Tra.) error are computed based on Three-point label, i.e., $a_1$, $a_2$, $a_3$. The standard deviation is in parentheses, and the $\downarrow$ sign indicates smaller values being better. The best results are in bold.}\label{tab:dif_atlas}
\begin{tabular}{ccccccc}
\toprule
\multirow{2}{*}{Reg. Meth.} & \multirow{2}{*}{Backbone} & \multicolumn{5}{c}{Three-Point and corresponding Rot. and Tra.}                                                                            \\ \cmidrule{3-7}
                                        &                           & \multicolumn{1}{c}{$a_1$ ED (mm) $\downarrow$} & \multicolumn{1}{c}{$a_2$ ED (mm) $\downarrow$} & \multicolumn{1}{c}{$a_3$ ED (mm) $\downarrow$} & \multicolumn{1}{c}{Rot. GD (°) $\downarrow$} & Tra. ED (mm) $\downarrow$ \\ \cmidrule{1-7}
                                        & VGGNet19                    & \multicolumn{1}{c}{3.84 (5.38)}    & \multicolumn{1}{c}{6.62 (9.65)}  & \multicolumn{1}{c}{5.98 (9.49)}  & \multicolumn{1}{c}{4.78 (8.25)}    & 3.84 (5.38)     \\  
                                     & MobileNetV2               & \multicolumn{1}{c}{3.98 (5.07)}    & \multicolumn{1}{c}{5.31 (8.69)}  & \multicolumn{1}{c}{5.70 (8.09)}  & \multicolumn{1}{c}{4.25 (7.02)}    & 3.98 (5.07)     \\
SyN                                         & ResNet50               & \multicolumn{1}{c}{3.50 (4.37)}    & \multicolumn{1}{c}{3.66 (5.17)}  & \multicolumn{1}{c}{4.31 (6.51)}  & \multicolumn{1}{c}{3.05 (4.65)}    & 3.50 (4.37)    \\ 
                                        & ResNet18                  & \multicolumn{1}{c}{3.48 (4.06)}    & \multicolumn{1}{c}{3.64 (4.77)}  & \multicolumn{1}{c}{4.30 (5.72)}  & \multicolumn{1}{c}{3.02 (4.06)}    &  3.48 (4.06)    \\
                                        \cmidrule{2-7}
                                        & Swin-T               & \multicolumn{1}{c}{\textbf{3.11} (3.02)}    & \multicolumn{1}{c}{\textbf{3.24} (4.71)}  & \multicolumn{1}{c}{\textbf{3.98} (5.86)}  & \multicolumn{1}{c}{\textbf{2.67} (4.32)}    &  \textbf{3.11} (3.02)    \\  \cmidrule{1-7}
                                        & VGGNet19                    & \multicolumn{1}{c}{3.96 (4.68)}    & \multicolumn{1}{c}{7.72 (10.48)}  & \multicolumn{1}{c}{6.72 (10.65)}  & \multicolumn{1}{c}{4.84 (8.02)}    &  3.96 (4.68)   \\
                                   & MobileNetV2               & \multicolumn{1}{c}{4.08 (4.44)}    & \multicolumn{1}{c}{6.44 (9.52)}  & \multicolumn{1}{c}{6.44 (8.25)}  & \multicolumn{1}{c}{4.46 (6.81)}    &  4.08 (4.44)   \\   
Affine                                         & ResNet50               & \multicolumn{1}{c}{3.52 (3.66)}    & \multicolumn{1}{c}{4.79 (6.18)}  & \multicolumn{1}{c}{5.05 (6.27)}  & \multicolumn{1}{c}{3.26 (4.52)}    & 3.52 (3.66)    \\ 
                                        & ResNet18                  & \multicolumn{1}{c}{3.48 (3.36)}    & \multicolumn{1}{c}{4.76 (5.60)}  & \multicolumn{1}{c}{5.04 (5.88)}  & \multicolumn{1}{c}{3.23 (3.83)}    &  3.48 (3.36)    \\
                                        \cmidrule{2-7}
                                        & Swin-T               & \multicolumn{1}{c}{\textbf{3.14}  (3.39)}    & \multicolumn{1}{c}{\textbf{4.38} (5.53)}  & \multicolumn{1}{c}{\textbf{4.92} (6.02)}  & \multicolumn{1}{c}{\textbf{2.84} (4.49)}    & \textbf{3.14} (3.39)    \\  \cmidrule{1-7}
                                        & VGGNet19                    & \multicolumn{1}{c}{7.80 (7.44)}    & \multicolumn{1}{c}{11.04 (11.16)}  & \multicolumn{1}{c}{10.94 (12.05)}  & \multicolumn{1}{c}{9.34 (10.78)}    &  7.75 (7.44)    \\ 
                                   & MobileNetV2               & \multicolumn{1}{c}{7.56 (7.20)}    & \multicolumn{1}{c}{9.76 (10.20)}  & \multicolumn{1}{c}{10.66 (9.66)}  & \multicolumn{1}{c}{9.01 (9.70)}    & 7.56 (7.20)   \\ 
Rigid                                         & ResNet50               & \multicolumn{1}{c}{7.35 (6.22)}    & \multicolumn{1}{c}{9.11 (9.03)}  & \multicolumn{1}{c}{9.30 (9.89)}  & \multicolumn{1}{c}{8.70 (8.38)}    & 7.35 (6.22)    \\ 
                                         & ResNet18                  & \multicolumn{1}{c}{ 7.32 (6.12)}    & \multicolumn{1}{c}{ 9.08 (8.28)}  & \multicolumn{1}{c}{9.26 (9.27)}  & \multicolumn{1}{c}{8.68 (7.79)}    &  \textbf{7.32} (6.12)   \\ 
                                         \cmidrule{2-7}
                                        & Swin-T               & \multicolumn{1}{c}{\textbf{6.97} (6.25)}    & \multicolumn{1}{c}{\textbf{8.73} (8.21)}  & \multicolumn{1}{c}{\textbf{8.68} (9.41)}  & \multicolumn{1}{c}{\textbf{8.19} (8.05)}    & \textbf{6.97} (6.25)    \\  \bottomrule
\end{tabular}
\end{table*}

\subsubsection{External Generalization Validation}
\begin{figure}[t]
\centering{\includegraphics[width=\columnwidth]{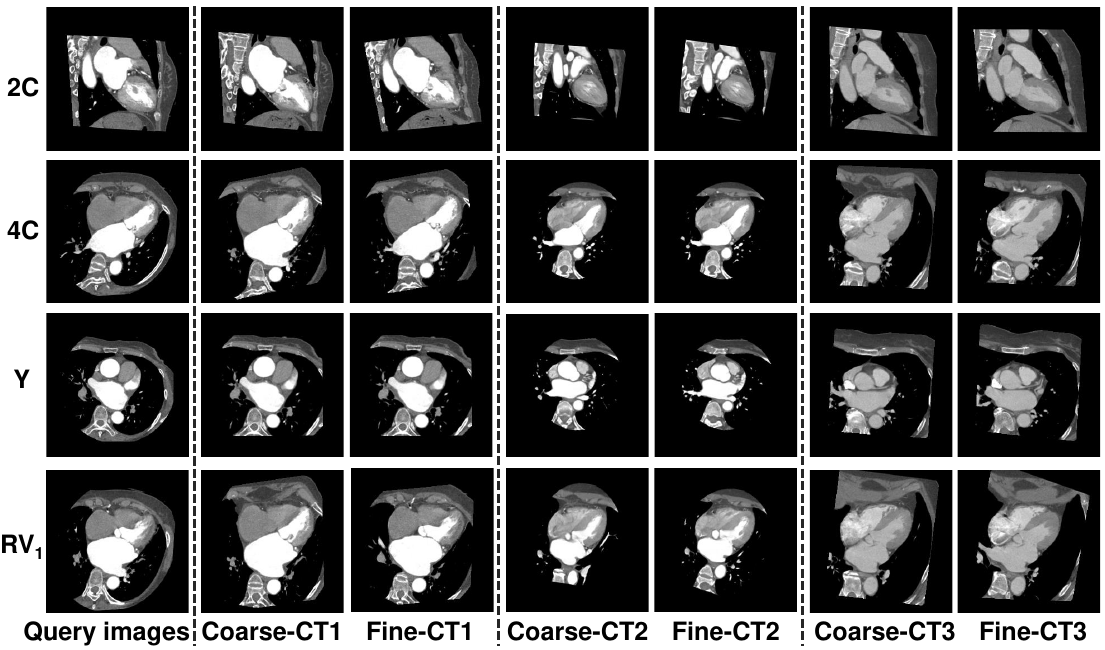}}
\caption{External generalization validation: visualization of positioning results on three different target CT volumes (CT1, CT2, CT3). The first column is four query images, resampling from CT1. The coarse and fine positioning results on CT1 are shown in the second and third columns, while the positioning results on CT2 are displayed in the fourth and fifth columns, and the sixth and seventh columns present the results for CT3.
}\label{fig:vis_res_mmwhs}
\end{figure}

To further validate the generalization ability of our proposed method, we similarly select four query images (2C, 4C, Y, and RV$_1$) and conduct experiments on 20 target CT volumes from \textit{Dataset 2}. Table \ref{tab:our-mmwhs} also shows the average quantitative results of the optimization-based method Opt-SVR, the learning-based method Lea-AVP, and our proposed framework AVP-AP. As we can see, our method outperforms the traditional Opt-SVR, and the SSIM metric is 3\% better than Lea-AVP on the 4C plane. Furthermore, Fig. \ref{fig:vis_res_mmwhs} visualizes the positioning results in three different CT volumes (CT1, CT2, CT3), where the query images are resampled from CT1. We can see that although the query images originate from CT1, our framework AVP-AP can effectively position similar slices in CT1, CT2, and CT3. This indicates that our framework has better generalization capability, and the query images can come from any different CT volume.  

\subsection{Ablation Study}
\label{sec:ablation}
We perform ablation experiments on the atlas prompting in terms of different registration methods, network architectures, and loss functions, so as to find optimal parameters for view positioning in atlas prompts.

\subsubsection{Effect of Different Registration Methods}
To evaluate the impact of different registration methods on PosNet in the atlas prompting, we train networks on data with Three-Point label generated by three different registration methods: SyN, Affine, and Rigid. Table \ref{tab:dif_atlas} shows the quantitative metrics' mean error and standard deviation on the test set. Here, $a_1$, $a_2$, and $a_3$ are the three points of the slices, while $Rot.$ and $Tra.$ represent the corresponding rotations and translations calculated based on $a_1$, $a_2$, and $a_3$. Then, we can see that the SyN registration achieves the most minimum average errors across all metrics, followed by the affine registration, with the rigid registration being the worst. This shows that good alignment is crucial for the automatic mapping of slices in the atlas space by training PosNet. Moreover, under the same ResNet18 network architecture, the results show that the rotation error of affine registration is only 0.21$^{\circ}$ larger than that of SyN, and the translation error is the same. This indicates that both registration methods have comparable alignment and an equivalent impact on PosNet's performance. In contrast, the rotation error of the rigid registration is nearly 5.66$^{\circ}$ larger, and the translation error is nearly 3.84 mm larger than those of SyN and affine registration under ResNet18. This suggests that rigid registration does not align the cardiac structures well, thereby affecting the performance of PosNet.

\subsubsection{Effect of Different Network Architectures}
\begin{figure}[t]
\centering
\begin{minipage}{0.156\linewidth}
    \centering
    \includegraphics[width=\textwidth]{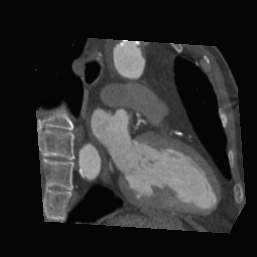}
\end{minipage}
\hfill
\begin{minipage}{0.156\linewidth}
    \centering
    \includegraphics[width=\textwidth]{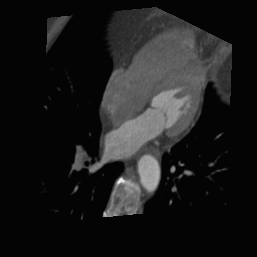}
\end{minipage}
\hfill
\begin{minipage}{0.156\linewidth}
    \centering
    \includegraphics[width=\textwidth]{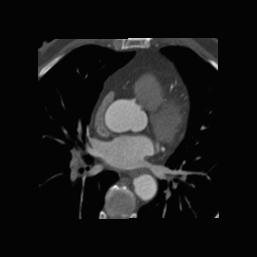}
\end{minipage}
\hfill
\begin{minipage}{0.156\linewidth}
    \centering
    \includegraphics[width=\textwidth]{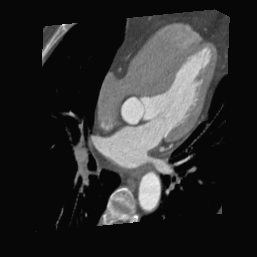}
\end{minipage}
\hfill
\begin{minipage}{0.156\linewidth}
    \centering
    \includegraphics[width=\textwidth]{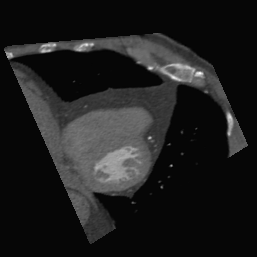}
\end{minipage}
\hfill
\begin{minipage}{0.156\linewidth}
    \centering
    \includegraphics[width=\textwidth]{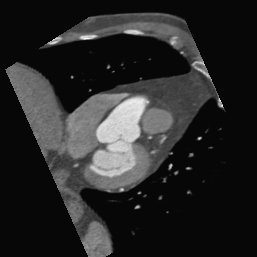}
\end{minipage}

\vspace{2pt} % Adjust the vertical space between the rows

\begin{minipage}{0.156\linewidth}
    \centering
    \includegraphics[width=\textwidth]{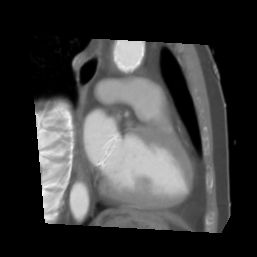}
    \centerline{(a) 2C}
\end{minipage}
\hfill
\begin{minipage}{0.156\linewidth}
    \centering
    \includegraphics[width=\textwidth]{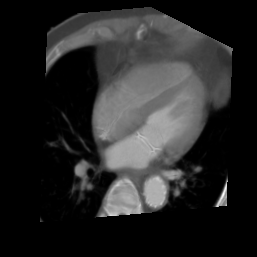}
    \centerline{(b) 4C}
\end{minipage}
\hfill
\begin{minipage}{0.156\linewidth}
    \centering
    \includegraphics[width=\textwidth]{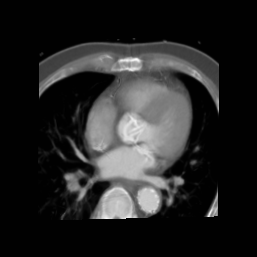}
    \centerline{(c) Y}
\end{minipage}
\hfill
\begin{minipage}{0.156\linewidth}
    \centering
    \includegraphics[width=\textwidth]{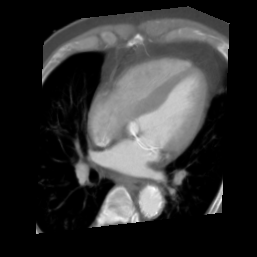}
    \centerline{(d) RV$_1$}
\end{minipage}
\hfill
\begin{minipage}{0.156\linewidth}
    \centering
    \includegraphics[width=\textwidth]{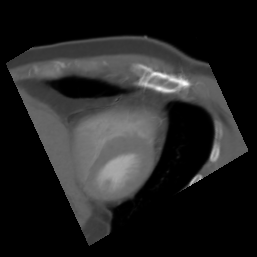}
    \centerline{(e) RV$_2$}
\end{minipage}
\hfill
\begin{minipage}{0.156\linewidth}
    \centering
    \includegraphics[width=\textwidth]{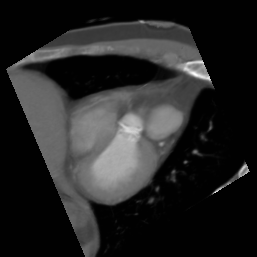}
    \centerline{(f) RV$_3$}
\end{minipage}
\caption{(a)-(f) represent six different cardiac CT slice views. Top row: six different query images. Bottom row: slices resampled from the SyN atlas volume using the positions predicted by PosNet.}\label{fig:atlas_res}
\end{figure}

As shown in Table \ref{tab:dif_atlas}, we assess the effect of five different network architectures on PosNet for each registration method. It can be seen that the transformer-based Swin-T outperforms all CNN-based backbones, with 0.35$^{\circ}$ and 0.37mm smaller rotation and translation errors than the best CNN-based ResNet18 in the SyN registration. This may be because global information is crucial for locating 2D slices within 3D space. Moreover, the self-attention mechanism with shifted windows in Swin-T can efficiently capture global and local information, and its hierarchical design facilitates multi-scale feature extraction. As a result, it exhibits a stronger feature representation capability, achieving superior performance. In addition, among the CNN-based backbones, ResNet18 achieves the lowest rotation error and translation error in all registration methods. Specifically, in the SyN registration, the mean rotation and translation errors of ResNet18 are smaller than those of ResNet50, MobileNetV2, and VGGNet19 by 0.03$^{\circ}$, 1.23$^{\circ}$, 1.76$^{\circ}$ and 0.02 mm, 0.50 mm, 0.36mm. We can observe that the errors of ResNet18 and ResNet50 are similar, with ResNet18 performing slightly better but within the error margin. This indicates that simply increasing the number of layers does not reduce the error, whereas switching to the transformer-based architecture resulted in lower errors. This further emphasizes the importance of global information for locating 2D slices in 3D space.
Fig. \ref{fig:atlas_res} presents examples of predictions made by the optimal PosNet. The first row is the given six query images, and the second row is 2D slices resampled from the SyN atlas volume using the positions predicted by PosNet. From the slices shown in all columns, PosNet can distinguish critical anatomical structures and perform precise slice positioning.

\subsubsection{Effect of Different Loss Functions/Label Parameterizations}\label{sec:loss}
Additionally, we evaluate the effect of different loss functions, i.e., label parameterization, on the accuracy of the PosNet. Here, we utilize the data generated by the affine registration methods and choose the same ResNet18 architecture. Table \ref{tab:dif_loss} presents the quantitative evaluation results of models trained with two different loss functions. As we can see, the results of the Three-Point label outperform those of the Rotvec-Cartesian label, except for a slightly higher average rotation error. This indicates that the loss function combination of the Three-Point label in the same Euclidean space is balanced and can improve the model's accuracy. Moreover, the average rotation error of the Rotvec-Cartesian label is slightly better than that of the Three-Point label, which may indicate the geodesic distance can accurately estimate the distance between two rotations, thus making the training results better. In addition, we can also observe that the training time for the Three-Point label is slightly shorter than that of the Rotvec-Cartesian label, possibly because the computation of geodesic distance involves matrix decomposition, which takes more time than the calculation of mean square error. 

\begin{figure}[t]
\centering
\begin{minipage}{0.45\linewidth}
    \vspace{6pt}
    \centerline{\includegraphics[width=\textwidth]{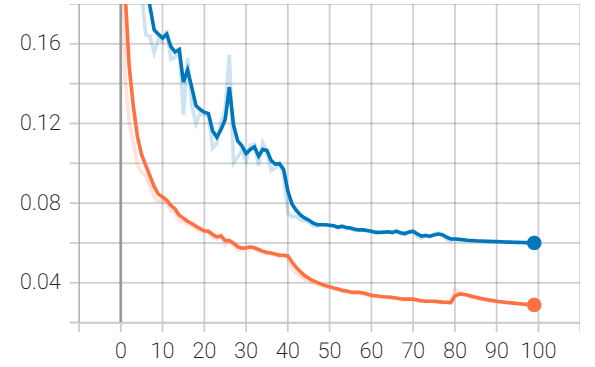}} 
    \centerline{(a) Rotvec-Cartesian label}
\end{minipage}
\begin{minipage}{0.45\linewidth}
    \vspace{3pt}
    \centerline{\includegraphics[width=\textwidth]{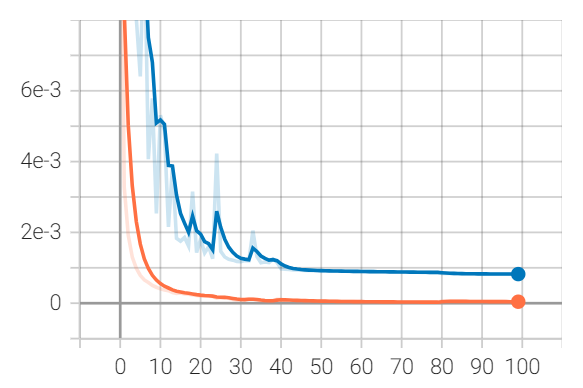}}
    \centerline{(b) Three-Point label}
\end{minipage}

\caption{Loss curves of training and validation for two different loss functions. (a) Loss curves for training with the Rotvec-Cartesian label; (b)  Loss curves for training with the Three-Point label. The orange line represents the training loss, and the blue line denotes the validation loss.}\label{fig:loss_line}
\end{figure}

Fig. \ref{fig:loss_line} shows the training and validation loss curves for two different loss functions. We can see that the loss curve of the Three-Point converges faster than that of Rotvec-Cartesian. Also, the training loss and validation loss for the Three-Point label have converged smoothly, with only a minor difference. In contrast, the training loss and validation loss for the Rotvec-Cartesian label converge more slowly, and the discrepancy between them is relatively large. The possible reason is that the Three-Point loss function $Loss_{tp}$ is a balanced combination within the same Euclidean space. In contrast, the Rotvec-Cartesian $Loss_{lc}$ combines SO(3) and Euclidean space, leading to an imbalance. In conclusion, both loss functions for the two representations are applicable, but the loss function of the Three-Point label performs better than that of Rotvec-Cartesian label.

\begin{table}[t]
\centering
\caption{Mean error and standard deviation of PosNet quantitative metrics for different loss functions. The standard deviation is in parentheses. The best results are highlighted in bold.}\label{tab:dif_loss}
\begin{tabular}{cccc}
\toprule
                 & Rot. GD (°) $\downarrow$ & Tra. ED (mm) $\downarrow$ & Training Time $\downarrow$\\ \cmidrule{1-4} 
Rotvec-Cartesian    & \textbf{3.09} (6.35) & 6.00 (4.32) & 2days 10hours   \\
Three-Point      & 3.23 \textbf{(3.83)} & \textbf{3.48 (3.36)} & \textbf{2days 8hours} \\ \bottomrule
\end{tabular}
\end{table}

% Discuss
\section{Discussion}
The human body has relatively fixed anatomical relationships, and the position of each structure serves as an index for the information of the structure. Accurate localization and navigation of each structure require bridging the semantic space and the pose space of the cross-sections of each structure. The experiment in Section \ref{sec:exis_real} validates that our proposed framework achieves automatic view positioning, completing the mapping from the query image to any 3D volume space. In addition, the comparison with existing work and radiologists further verifies that our proposed atlas prompting-based framework, which allows input of arbitrary views in a reference CT, performs better in multi-view positioning tasks. Although our work focuses on cardiac CT, our framework is actually independent of the organ and modality. In the future, this approach can be extended to various scenarios involving the heart and other organs, such as cardiac ultrasound navigation, enabling full-organ navigation.

Regarding atlas-guided prompts, our current input is limited to images. However, with the rapid development of language models and multimodal large language models, we can achieve semantic alignment between natural languages, images, and voices in the future, allowing us to expand prompts to include text or voice. Elaborately, we can extract the corresponding description information of arbitrary slices from CT reports. Then, based on the above paired data, we can align data from two modalities in the feature space using the CLIP model. Training multimodal data in this way not only can enhance the model's performance, but also improve the functionality of atlas prompting. Overall, the input prompt can include both the standard or arbitrary views used in routine exams and detailed descriptions of specific structures or personalized features.
 
Concerning the generalizability of our proposed method, we have conducted internal testing on a private dataset and performed external generalization validation on a public dataset. Although our private dataset comes from a single hospital and is relatively limited in size, the promising results achieved on the public dataset indicate that our method exhibits good generalizability. This may be because our private dataset includes cardiac CT scans from both healthy individuals and patients with coronary artery disease, ensuring the diversity and richness of the dataset; moreover, data augmentation techniques such as brightness transformation were applied to simulate and expand the data under different imaging conditions, which further improved the generalization capability of our method. In the future, multi-center data from multiple hospitals could be collected to validate and adapt our framework for generalizability across different data sources, domains, and imaging qualities. On one hand, multi-center data can be used to develop the critical atlas prompting technique, enabling it to generalize well across diverse data sources, domains, and imaging qualities; On the other hand, the existing model can be further validated on data from other centers. Additionally, more sophisticated data augmentation techniques, such as artifact simulation and noise addition, can be employed to simulate imaging characteristics from different sources, domains, and imaging qualities, further improving the generalization capability of our approach. In addition, the main contribution of this paper is our AVP-AP framework, which is independent of the pre-trained models, and any available pre-trained model can be used. Here, we selected BiomedCLIP with strong anatomical feature extraction capability for our research. As for alternative feature extraction methods, there are already well-established pre-trained modes for medical images, such as MedCLIP and RadFM, as well as pre-trained models for natural images. Furthermore, we can also train our own pre-trained model using existing data for feature extraction. As depicted in Fig. \ref{fig:task}, our framework is also flexible and capable of positioning arbitrary slices without retraining. In the future, deep learning-based registration can achieve real-time performance, and we will attempt to use deep learning-based registration. Then, we can perform view positioning using the faster algorithm, and ultimately, achieve lightweight deployment for real-time applications in practical scenarios.

% 5 Conclusion
\section{Conclusion}
In this paper, we propose a method for automatic view positioning, designed to identify the most similar target 2D slices and their corresponding positions within the target 3D CT volume by utilizing our atlas prompting and iterative similarity search based on a given foundation model. We design a coarse-to-fine framework AVP-AP, where the coarse network, by aligning with a unified atlas of cardiac CT, retrieves corresponding positions in the atlas space from the atlas prompting process. These positions are then aligned into the target 3D CT volume space through rigid registration to further generate a series of coordinated candidates and corresponding 2D slices. In the refinement stage, we obtain accurate 2D slices and precise positions by filtering the foundation model-based image features similarity of the candidate image slices captured from the coarse stage. Experimental results show that our method is better than the existing work and radiologists in terms of accuracy and efficiency for view positioning. In addition, experiments on a public dataset verified that our method has good generalization ability. Hence, our approach is of significant importance for clinical view positioning in many scenarios. For example, atlas prompts can facilitate radiologists to locate and review images rapidly, and surgeons can efficiently utilize historical data to search for similar successful cases for better surgical planning.

\bibliographystyle{IEEEtran}
\bibliography{IEEEabrv,reference}

\end{document}